\documentclass[aps,prc,showpacs,twocolumn,amssymb,floatfix]{revtex4}

\usepackage{graphicx}

\begin{document}

\title{Transverse  and longitudinal 
momentum spectra of fermions \\
produced in strong SU(2) fields}
\author{Vladimir V. Skokov}
\affiliation{Bogoliubov Laboratory of Theoretical Physics, 
Joint Institute for Nuclear Research, \\
141980, Dubna, Russia}
\author{P\'eter L\'evai}
\affiliation{RMKI Research Institute for Particle and Nuclear Physics, \\
P.O. Box 49, Budapest 1525, Hungary}

\date{14 August  2008}

\begin{abstract}
We study the transverse and longitudinal momentum spectra of fermions 
produced in a strong, time-dependent non-Abelian SU(2) field.
Different time-dependent field strengths are introduced. The
momentum spectra are calculated for the produced 
fermion pairs in a kinetic model. The obtained spectra 
are similar to the Abelian case, and they
display exponential or polynomial behaviour at high $p_T$,
depending on the given time dependence.
We investigated different color initial conditions and discuss
the recognized scaling properties for both Abelian and SU(2) cases.
\end{abstract}


\maketitle

 \section{Introduction}
During last years large amount of data on particle spectra have been 
collected in relativistic heavy ion collisions 
at the Super-Proton Synchrotron (SPS, CERN) and at
the Relativistic Heavy Ion Collider (RHIC, BNL) at
c.m. energy of $\sqrt{s} = 10 - 200$ GeV~\cite{QM04,QM05}
in a wide transverse momentum range, 
$0 \leq p_T \leq 20$ GeV. 
Since the microscopic mechanisms of hadron production in 
hadron-hadron and heavy ion collisions are not fully understood,
thus it is very important to improve our theoretical  
understanding on this field.  The forthcoming hadron
and heavy ion experiments at the Large Hadron Collider
(LHC, CERN) at $\sqrt{s} = 5500$ GeV
will increase the transverse momentum window to $0 \leq p_T \leq 50$ GeV.
Thus LHC experiments will become a decisive test
between different perturbative and non-perturbative models of
hadron formation, especially in the high-$p_T$ region.

Theoretical descriptions of particle production in high energy $pp$ 
collisions are based on the introduction of chromoelectric flux tube 
('string') models, where these tubes are connecting the quark and 
diquark constituents of the colliding 
protons~\cite{FRITIOF,HIJ,RQMD,Topor0507}. 
String picture is a good example of how to convert the kinetic
energy of a collision into field energy. New hadrons will be produced
via quark-antiquark and diquark-antidiquark pair production from the
field energy, namely from the unstable flux tubes.  
These models can describe experimental data 
very successfully at small $p_T$, especially at $p_T < 2-3$ GeV. 
At higher $p_T$ one can apply perturbative QCD-based 
models~\cite{FieldFeyn,Wang00,Yi02}, which can provide the necessary
precision to analyse nuclear effects in the nuclear collisions.

However, at RHIC and LHC energies the string density is expected to be 
so large that a strong collective gluon field will be formed in the whole
available transverse volume.
Furthermore, the gluon number will be so high that a classical gluon field
as the expectation value of the quantum field can be considered
and investigated in the reaction volume.
The properties of such non-Abelian classical fields and details
of gluon production 
were studied very intensively during the last years, especially
asymptotic solutions (for summaries see Refs.~\cite{McL02,IanVen03}).
Fermion production was calculated recently~\cite{Gelis04,Blaiz04}.
Lattice calculations were performed to describe 
strong classical fields under finite space-time 
conditions in the very early stage of heavy ion collision,
especially immediately after the overlap~\cite{Krasnitz,Lappi,Lappi_more}. 
New methods have been developed to investigate 
the influence of inhomogeneity on particle 
production~\cite{Gies05,Dunne05,Kim06}.

Fermion pair production together with boson pair production
were investigated by different kinetic models of particle
production from strong 
Abelian~\cite{Gatoff87,Kluger91,Gatoff92,Wong95,Eisenberg95,Vin02,Per03,Per06}
and non-Abelian~\cite{Prozor03,Diet03,Nayak} fields. 
These calculations concentrated
mostly on the bulk properties of the gluon and quark matter, 
the time evolution of the system,
the time dependence of energy and particle number densities, 
and the appearance of fast thermalization.

Our main interest is the transverse momentum distribution of
produced fermions and bosons, however the longitudinal momentum
can become equally interesting considering wide rapidity ranges. 
In our previous paper (see Ref.~\cite{SkokLev05})
we investigated the Abelian case, namely particle pair-production in
a strong external electric field.
We have focused on the numerical solution of a kinetic model
and discussed the influence of the field strength applying
different time dependences.
We have demonstrated a 
scaling behaviour in time and transverse momenta, namely
$t\cdot E_0^{1/2}$ and $k_T/E_0^{1/2}$.
The kinetic equation and the numerical calculation 
yielded a fermion dominance in the mid-rapidity region.
In case of realistic Bjorken type time evolution our
numerical result on fermion spectra has overlapped the boson spectra
obtained in 1+2 dimensional lattice calculations both in magnitude
and shape. 

In this paper we solve the kinetic
model in the presence of an SU(2) non-Abelian color field. We
focus on the determination of the appropriate kinetic equation system
and its numerical solution in case of different initial conditions.
Section 2 summarizes the kinetic equation for color Wigner function.
In Section 3 we introduce phenomenologically the fermion distribution function.
In Section 4 the kinetic equation and its simplifications are 
displayed in detail, especially with zero fermion mass.
In Section 5 we present kinetic equation for the case, when external non-Abelian 
field have fixed direction in color space. 
The appropriate
form of the kinetic equation is determined, which was solved numerically
at different time-dependent strong fields. 
Results on field components and fermion distribution functions 
are displayed in Section 6.
We display recognized scaling properties of the solutions in Section 7.
In Appendix A we show a special non-Abelian setup, which
leads to the similar results obtained in the Abelian case.

\section{The kinetic equation for the Wigner function}

The fermion production in a strong external field can be characterized by
a space homogeneous Wigner function, $W(\mathbf{k}; t)$.
The evolution of this Wigner function 
is investigated by the kinetic equation
in the frame of the covariant single-time 
formalism, where a time-dependent 
Abelian ($A^\mu$)~\cite{BialBirul91,Heinz96,Heinz98,Holl02}
or non-Abelian ($A_\mu^a$)~\cite{Heinz98,Prozor03}
external field is included.
Here we choose a longitudinally dominant 
color vector field in Hamilton gauge described by
the 4-potential
\begin{equation}
A_\mu^a = (0,-{\bf{A}}^a) = (0,0,0,A_3^a) \ .
\end{equation}
The Wigner function depends on the 3-momenta $\mathbf{k}=(k_1,k_2,k_3)$.  
In the "instant" frame of 
reference~\cite{Prozor03} we obtain the 
kinetic equation for $W(\mathbf{k}; t)$ as
\begin{eqnarray}
&&\partial_t W+ \frac{g}{8}\frac{\partial}{\partial k_i}
\left( 4\{W,F_{0i}\}+ \right. \nonumber \\
&&\ \ \ \ \ \ \ \left. + 
2\left\{F_{i\nu},[W,\gamma^0 \gamma^\nu]\right\}-
\left[F_{i\nu},\{W,\gamma^0 \gamma^\nu\}\right] \right)=\nonumber\\
&&=ik_i \{\gamma^0 \gamma^i,W\}-im[\gamma^0,W] +ig \left[A_i\,, 
[\gamma^0 \gamma^i ,W]\right] . \ \ \ 
\label{wigner_gen}
\end{eqnarray}
Here $m$ denotes the current mass of the fermion 
produced in the strong field and $g$ is the coupling constant.

The color decomposition
with $SU(N_c)$ generators in fundamental representation 
($t^a$) is
\begin{equation}
W = W^s + W^a t^a, \,\, \ \ \ a = 1,2,..., N_c^2-1 \ ,
\label{color_decomposition}
\end{equation}
where $W^s$ is the color singlet and $W^a$ is the color
multiplet component (triplet in $SU(2)$ with $N_c=2$).

The spinor decomposition is the following:
\begin{equation}
W^{s|a} = a^{s|a} + b^{s|a}_\mu \gamma^\mu + c^{s|a}_{\mu\nu} \sigma^{\mu\nu} +
d^{s|a}_\mu \gamma^\mu \gamma^5 + i e^{s|a} \gamma^5.
\label{Clifford_decomposition}
\end{equation}

The time evolution of the Wigner function is described by 
eq.~(\ref{wigner_gen}). Using the color and spinor
decomposition terms in 
eqs.(\ref{color_decomposition})-(\ref{Clifford_decomposition}), 
we will determine the time evolution
numerically. However, our main goal is to obtain the transverse
momentum spectra of the produced fermion pairs, thus we need to
introduce the fermion distribution function.
There is no straightforward way to define this distribution 
function, but we can find a phenomenological way, summarized in the
next chapter.

\section{The fermion distribution function from the energy density} 

The energy density carried by the produced fermions is  defined 
throughout the Wigner function~\cite{Heinz98,Prozor03}:
\begin{equation}
\varepsilon_f  = \mbox{Tr} \langle \, (m-\gamma^i k_i) W  + \omega(\mathbf{k})\, \rangle \ ,
\label{energy}
\end{equation}
where the second term fixes the zero energy level to the physical value (note, that expression (\ref{energy}) 
is free from divergent vacuum contribution).  
Here the one particle energy is denoted by 
$\omega(\mathbf{k}) = \sqrt{ \mathbf{k}^2 + m^2}$.
The trace is over color and spinor and the averaging is taken 
over momenta:
\begin{equation}
\langle \, {\cal X} \, \rangle = 
\int \frac{d^3k}{(2 \pi)^3}  {\cal X}(\mathbf{k}; t).
\label{averaging}
\end{equation}
After the  color and spinor decomposition  one obtains
\begin{equation}
\varepsilon_f  = 2 N_c \langle \,  2 m a^s + 2 \mathbf{k} \, \mathbf{b}^s  
+  \omega(\mathbf{k}) \, \rangle. 
\label{energy_after_decompos}
\end{equation}
In parallel, focusing on one-particle distribution function
the usual energy density formula has the following form:
\begin{equation}
\varepsilon_f  = 
4 N_c \langle \, \omega(\mathbf{k}) f_f(\mathbf{k}) \, \rangle = 
4 N_c  \int \frac{d^3k}{(2 \pi)^3} 
\omega(\mathbf{k}) f_f(\mathbf{k}).
\label{usual}
\end{equation}
Note, that both averages (\ref{energy}) and (\ref{usual}) may contain ultraviolet 
divergence, that can be properly regularized (for Abelian fields see e.g.~\cite{Cooper:1992hw,grib}). 
However, since we are not going to calculate bulk properties of produced pairs, 
the issue of regularization is not important  in current consideration and will be considered elsewhere.

Now, combining eq.(\ref{energy_after_decompos}) 
and eq.(\ref{usual}) one can derive
phenomenologically a distribution function, namely
\begin{equation}
f_f(\mathbf{k},t) = \frac{m a^s(\mathbf{k},t) +  \mathbf{k} \, \mathbf{b}^s(\mathbf{k},t)}
{\omega(\mathbf{k})}  + \frac{1}{2}\ .
\label{DF}
\end{equation}
The time dependence in $f_f(\mathbf{k},t)$
is connected to the time-evolution of 
$a^s(\mathbf{k},t)$ and $\mathbf{b}^s(\mathbf{k},t)$, which
is followed by solving the decomposed 
kinetic equation of eq.~(\ref{wigner_gen}).

The fermion distribution function should be zero in vacuum.
The expression in eq.~(\ref{DF}) satisfies this request. 
Indeed 
the vacuum solution ($A_\mu = 0$) for the singlet Wigner function has the 
following form (see Ref.~\cite{Prozor03})
\begin{equation}
W^{s} = - \frac{1}{2} \frac{m+\mathbf{k} 
\mathbf{\gamma}}{\omega(\mathbf{k})} \ .
\label{WFvac}
\end{equation}
The vacuum solution for the multiplet Wigner function is zero,
$W^a = 0$. The substitution of this Wigner function
into the definition of the distribution function in eq.(\ref{DF})
leads to $f_f=0$ in  vacuum, which is physically correct.
Furthermore, one can see that $f_f(\mathbf{k},t)$ 
is positive definite 
 in presence of a non-zero field ($A_\mu \ne 0$).

Although our fermion distribution
function $f_f(\mathbf{k},t)$ has been postulated phenomenologically and 
possibly it is not a unique solution, but it seems to be the right tool
to investigate the time evolution of the energy
distribution at the microscopical level in a space homogeneous 
environment.

\section{Kinetic equation for the Wigner function in SU(2)}

Our basic aim is to determine the fermionic distribution 
function and its time evolution in the presence of a 
strong, time-dependent non-Abelian  SU(2) field. 
For this task we need to substitute the color and spinor
decomposed Wigner function of eq.~(\ref{color_decomposition}) into
the kinetic equation in eq.~(\ref{wigner_gen}). 
After evaluating the color and spinor indeces we obtain
a system of coupled differential equations, 
which consists of 32 components in SU(2):
\begin{eqnarray}\label{hom_corrected}  
\partial_t a^s+\frac{g}{4}E^a\frac{\partial}{\partial k_3} a^a 
     &=& -4\mathbf{k} \mathbf{c}_1^s,\,\\
\partial_t a^a+gE^a\frac{\partial}{\partial k_3} a^s        
     &=& -4\mathbf{k}\mathbf{c}_1^a,\,\\
\partial_t b^s_0+\frac{g}{4}E^a\frac{\partial}{\partial k_3} b^a_0   
        &=&  0,\\ 
\partial_t b^a_0+g E^a\frac{\partial}{\partial k_3} b^s_0   
        &=& 2  g f^{abc}\mathbf{A}^b \mathbf{b}^c,\,\\
\partial_t \mathbf{b}^s +\frac{g}{4}E^a\frac{\partial}{\partial k_3} 
     \mathbf{b}^a &=& 2\mathbf{k}\times\mathbf{d}^s + 4 m \mathbf{c}^s_1,\,\\
\partial_t \mathbf{b}^a+gE^a\frac{\partial}{\partial k_3} \mathbf{b}^s  
     &=& 2\mathbf{k}\times\mathbf{d}^a +  4 m \mathbf{c}^a_1 + \nonumber \\
  && + 2 g  f^{abc}\mathbf{A}^b b_0^c  ,\,\\
\partial_t \mathbf{c}_1^s +\frac{g}{4}E^a\frac{\partial}{\partial k_3}
     \mathbf{c}_1^a &=& \mathbf{k} a^s - m \mathbf{b}^s,\,\\
\partial_t\mathbf{c}_1^a+gE^a\frac{\partial}{\partial k_3} \mathbf{c}_1^s  
      &=& \mathbf{k} a^a - m \mathbf{b}^a + \nonumber  \\ 
			&& -  2 g f^{abc}(\mathbf{A}^b\times \mathbf{c}_2^c),\,\\
\partial_t \mathbf{c}_2^s+\frac{g}{4}E^a\frac{\partial}{\partial k_3} 
        \mathbf{c}_2^a &=& \mathbf{k} e^s,\,\\
\partial_t\mathbf{c}_2^a+ gE^a\frac{\partial}{\partial k_3} \mathbf{c}_2^s
        &=& \mathbf{k} e^a -
2 g f^{abc}(\mathbf{A}^b\times \mathbf{c}_1^c),\,\\
\partial_t d^s_0+\frac{g}{4}E^a\frac{\partial}{\partial k_3} d^a_0  
        &=&  2 m e^s,\\ 
\partial_t d^a_0+g E^a\frac{\partial}{\partial k_3} d^s_0   
        &=&  2 m e^s + 2 g f^{abc}\mathbf{A}^b \mathbf{d}^c\, ,\\
\partial_t \mathbf{d}^s+\frac{g}{4}E^a\frac{\partial}{\partial k_3} 
 \mathbf{d}^a  &=& 2\mathbf{k}\times\mathbf{b}^s,\,\\
\partial_t \mathbf{d}^a+ gE^a\frac{\partial}{\partial k_3} \mathbf{d}^s
        &=& 2\mathbf{k}\times\mathbf{b}^a + 2 g  f^{abc}\mathbf{A}^b d_0^c ,\,\\
\partial_t e^s+\frac{g}{4}E^a\frac{\partial}{\partial k_3} e^a
        &=& -4\mathbf{k}\mathbf{c}_2^s  - 2 m d_0^s,\,\\
\partial_t e^a+\frac{g}{4}E^a\frac{\partial}{\partial k_3} e^s 
        &=& -4\mathbf{k}\mathbf{c}_2^a - 2 m d_0^a.
\end{eqnarray}
Here we use the notation
$b^{s|a}_\mu = (b^{s|a}_0, -\mathbf{b}^{s|a})$, and a similar one
for $d^{s|a}_\mu$.
The antisymmetric tensor component of the Wigner function
is defined as $c^{\mu\nu} = (\mathbf{c}_1, \mathbf{c}_2)$~\cite{Landau2}.
The strength of the color electric field is
$E^a \equiv E^a_3 = - \dot{A}_3^a$.


Since the mass of light quarks is small\footnote{Quark mass is the proper scale for static fields as 
we use in the case of Schwinger formula. For time dependent fields the inverse characteristic
time $\delta$ will substitute the mass term \cite{grib}, namely $\widehat{m} \propto 1/\delta$.
Our $1/\delta$ value (see Section IV) will determine particle production for zero quark mass 
in contrast to the Schwinger  formula.}, we neglect the mass term
in our SU(2) calculation.
In this case the distribution function for massless fermions is  
solely  defined  by $\mathbf{b}^s$: 
\begin{equation}
f_f(\mathbf{k},t) = \frac{\mathbf{k} \, 
\mathbf{b}^s} {|\mathbf{k}|} + \frac{1}{2} \ ,
\label{masslesDF}
\end{equation}

The zero fermion mass leads to the simplification of the above 
kinetic equation, which finally is splitted into two independent parts:
one for $a^{s|a}, \mathbf{c}_1^{s|a}, \mathbf{c}_2^{s|a}, e^{s|a}$
and another one for $b_0^{s|a}, \mathbf{b}^{s|a}, d_0^{s|a},
\mathbf{d}^{s|a}$. 
 One can recognize that the second set completely defines 
the evolution of the fermion distribution function.
This way the chiral symmetry of the massless case is manifested itself.
Thus we focus on these equations, only:
\begin{eqnarray}
\label{partII_1}
\partial_t \mathbf{b}^s +\frac{g}{4}E^a\frac{\partial}{\partial k_3} 
\mathbf{b}^a &=& 2\mathbf{k}\times\mathbf{d}^s,\,\\
\label{partII_2}
\partial_t \mathbf{b}^a+gE^a\frac{\partial}{\partial k_3} 
\mathbf{b}^s            &=& 2\mathbf{k}\times\mathbf{d}^a 
+2  g  f^{abc}\mathbf{A}^b b_0^c,\,\\
\label{partII_3}
\partial_t \mathbf{d}^s+\frac{g}{4}E^a\frac{\partial}{\partial k_3} 
\mathbf{d}^a  &=& 2\mathbf{k}\times\mathbf{b}^s,\,\\
\label{partII_4}
\partial_t \mathbf{d}^a+  gE^a\frac{\partial}{\partial k_3} 
\mathbf{d}^s 
&=& 2\mathbf{k}\times\mathbf{b}^a + 2 g  f^{abc}\mathbf{A}^b d_0^c,\,\\
\label{partII_5}
\partial_t b^s_0+\frac{g}{4}E^a\frac{\partial}{\partial k_3} 
b^a_0            &=&  0,\\ 
\partial_t b^a_0+g E^a\frac{\partial}{\partial k_3} b^s_0   
         &=&  2 g f^{abc}\mathbf{A}^b \mathbf{b}^c,\,\\
\partial_t d^s_0+\frac{g}{4}E^a\frac{\partial}{\partial k_3} d^a_0    
        &=&  0,\\ 
\partial_t d^a_0+g E^a\frac{\partial}{\partial k_3} d^s_0    
        &=& 2  g f^{abc}\mathbf{A}^b \mathbf{d}^c.
\label{partII_6}
\end{eqnarray}
The vacuum initial conditions are given by eq.(\ref{WFvac}):
\begin{eqnarray}
\mathbf{b}^s(t\to-\infty) = -\frac{\mathbf{k} }{2 |\mathbf{k}|}, 
\label{initials}
\end{eqnarray}
The other components of the Wigner function have
zero initial values.

 The system of equations in (\ref{partII_1})-(\ref{partII_6})
can be solved numerically for different color configurations.
Thus the evolution of the color could be followed in detail,
similarly to the color evolution 
in the Wong Yang-Mills equations~\cite{WongYM,Colorequi}. 
However, in the next section for the sake of simplicity we will consider assumption 
that the external field has a fixed direction in color space. 


\section{Kinetic equation in SU(2) with external field of fixed direction in color space}

The equations (\ref{partII_1})-(\ref{partII_6}) provides a full
description of the fermion production in a 
time-dependent SU(2) color field. Here we will consider the case of the external field,
that has a fixed direction in color space:  
$\mathbf{A}^a \equiv \mathbf{A}^\diamond n^a$, where $n^a n^a=1$ and $\partial_t n^a=0$,
in general case, and $A_3^a \equiv A^\diamond n^a$, $a=1,2,3$  in our calculation.
Then the equations (\ref{partII_1})-(\ref{partII_6}) have 
the following particular solution: 
\begin{equation}
\mathbf{b}^a = \mathbf{b}^\diamond n^a, \, \,\mathbf{d}^a = \mathbf{d}^\diamond n^a.  
\label{color_democ}
\end{equation}
Taking into account the zero initial conditions for $b_0^a, d_0^a$
and $b_0^s, d_0^s$,   
the equations (\ref{partII_5})-(\ref{partII_6}) become trivial
and do not contribute to the set of equations 
(\ref{partII_1}-\ref{partII_4}), the last simplifies to give 
\begin{eqnarray}
\label{partII_r1}
\partial_t \mathbf{b}^s +\frac{3g}{4}E^\diamond\frac{\partial}{\partial k_3} 
\mathbf{b}^\diamond &=& 2\mathbf{k}\times\mathbf{d}^s,\,\\
\label{partII_r2}
\partial_t \mathbf{b}^\diamond+gE^\diamond\frac{\partial}{\partial k_3} 
\mathbf{b}^s  &=& 2\mathbf{k}\times\mathbf{d}^\diamond, \,\\
\label{partII_r3}
\partial_t \mathbf{d}^s+\frac{3g}{4}E^\diamond\frac{\partial}{\partial k_3} 
\mathbf{d}^\diamond  &=& 2\mathbf{k}\times\mathbf{b}^s,\,\\
\label{partII_r4}
\partial_t \mathbf{d}^\diamond+ gE^\diamond\frac{\partial}{\partial k_3} 
\mathbf{d}^s      &=& 2\mathbf{k}\times\mathbf{b}^\diamond.
\end{eqnarray}
Introducing the unit vector collinear to the field direction,
$\mathbf{n} = \mathbf{E}^\diamond/\vert \mathbf{E}^\diamond \vert = (0,0,1)$,
and $\mathbf{k}_\perp = (k_1,k_2,0)$,
we can perform the following vector decomposition:
\begin{eqnarray}
\mathbf{b}^{\diamond|s} &=& b_3^{\diamond|s}  \mathbf{n} + 
b_\perp^{\diamond|s}  \frac{\mathbf{k}_\perp}{k_\perp}, \\
\mathbf{d}^{\diamond|s}  &=& d^{\diamond|s}  \mathbf{n} \times 
\frac{\mathbf{k}_\perp}{k_\perp}  \ .
\label{decomposition_axial_polar_g}
\end{eqnarray}
Finally we obtain 6 equations, only: 
\begin{eqnarray}
\partial_t b^s_\perp  + \frac{3 g}{4} E^\diamond \frac{\partial}{\partial k_3} 
b_\perp^\diamond &=& - 2 k_3 d^s \ ,
\label{final_massles_iso_beg} \\
\partial_t b^s_3  + \frac{3 g}{4} E^\diamond \frac{\partial}{\partial k_3} 
b_3^\diamond &=& 2 k_{\perp} d^s, \\ 
\partial_t d^s  + \frac{3 g}{4} E^\diamond \frac{\partial}{\partial k_3} d^\diamond 
&=& 2 k_3 b_{\perp}^s -2 k_{\perp} b_3^s   , \\ 
\partial_t b_\perp^\diamond  + g E^\diamond \frac{\partial}{\partial k_3} b^s_\perp 
&=& -2 k_3 d^\diamond, \\ 
\partial_t b_3^\diamond  + g E^\diamond \frac{\partial}{\partial k_3} 
b^s_3 &=& 2 k_{\perp} d^\diamond, \\ 
\partial_t d^\diamond  + g E^\diamond \frac{\partial}{\partial k_3} d^s 
&=&  2 k_{3} b_{\perp}^\diamond -2 k_{\perp} b_3^\diamond. 
\label{final_massles_iso_end}
\end{eqnarray}
These equations determine the time evolution
of the wanted distribution function:
\begin{equation}
f_f(\mathbf{k},t) =  
\frac{k_3 \, {b}^s_3 + k_\perp \, b^s_\perp} 
{\vert \mathbf{k} \vert}  + \frac{1}{2} \ .
\label{masslesDF_fin}
\end{equation}
The numerical solution of 
eqs.~(\ref{final_massles_iso_beg})-(\ref{final_massles_iso_end})
requires initial conditions. In our case the appropriate vacuum
initial conditions are the following:
\begin{eqnarray}
\label{initial_cond_beg}
b^s_{3}(\mathbf{k}, t=-\infty)  &=& -\frac{k_3}{2 k} \ ,  \\ 
b^s_{\perp}(\mathbf{k}, t=-\infty)  &=& -\frac{k_\perp}{2 k} \ , \\
b_{3}^\diamond (\mathbf{k}, t=-\infty)  &=& 
b_{\perp}^\diamond (\mathbf{k}, t=-\infty)= 0 \ , \\   
d^\diamond (\mathbf{k}, t=-\infty)&=&  
d^s(\mathbf{k}, t=-\infty) = 0.
\label{initial_cond}
\end{eqnarray}

Now we have all parts to proceed numerically for any time-dependent
field strength and determine the momentum distribution of the
produced fermions.

\section{Numerical results}

In heavy ion collisions, 
one can assume three different types of
time dependence
for the color field to be formed: 
a) pulse-like field develops with a fast increase, 
which is followed by a fast fall in the field strength;
b) formation of a constant field
($E_0$) is maintained after the fast increase in the initial time period;
c) scaled decrease of the field strength appears, 
which is caused by
particle production and/or transverse expansion, and the decrease
is elongated in time much further than the pulse-like assumption.

\begin{figure}[b]
\centerline{
\rotatebox{0}{\includegraphics[height=7.0truecm]
   {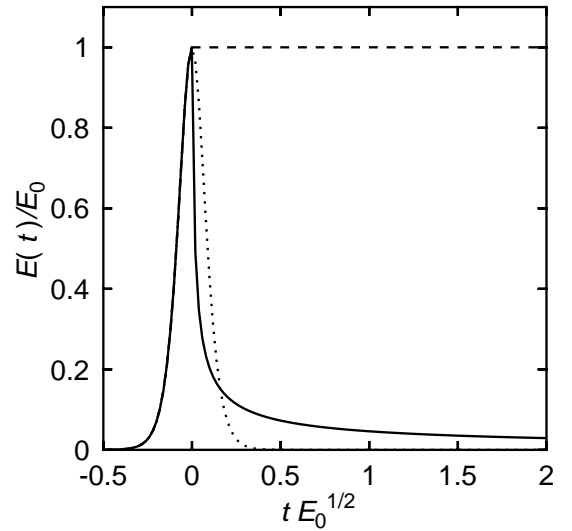}}
}
\caption{
The time dependence of
external field $E(t)$ in three physical scenarios:
a) pulse ({\it dotted line});
b) constant field, $E_0$ ({\it dashed line});
c) scaled decrease ({\it solid  line}).
}
\label{etime}
\end{figure}
Figure ~\ref{etime} displays three sets for the
time dependence of the external field~\cite{SkokLev05}:
\begin{eqnarray}
E_{pulse}^\diamond(t) &= E_0 \cdot \left[ 1- \textrm{tanh}^2 (t/\delta) \right] 
\hspace*{0.6truecm}   & \label{Ep} \\
E_{const}^\diamond(t) &= \left\{ \begin{array}{ll}
     E^\diamond_{pulse}(t) & {\textrm at} \ \ {t < 0} \\  
     E_0          & {\textrm at} \ \ {t \geq 0} 
                \end{array} \right. \label{Ec} \\
E_{scaled}^\diamond(t) &= \left\{ \begin{array}{ll}
     E^\diamond_{pulse}(t) & {\textrm at} \ \ {t < 0} \\ 
     \frac{E_0}{(1+t/t_0)^{\kappa}}          & {\textrm at} \ \ {t \geq 0} 
                \end{array} \right. \label{Es}
\end{eqnarray}

In eq.~(\ref{Ep}) we choose $\delta=0.1 / \, E_0^{1/2}$, 
which corresponds to RHIC energies.
In eq.~(\ref{Es}) the value
 $\kappa=2/3$ indicates a  longitudinally scaled Bjorken expansion
with $t_0 = 0.01 / \, E_0^{1/2}$. 
In fact, the whole time dependence
is scaled by ${E_0}^{1/2}$.

\begin{figure}[t]
\centerline{
\rotatebox{0}{\includegraphics[height=7.0truecm]
   {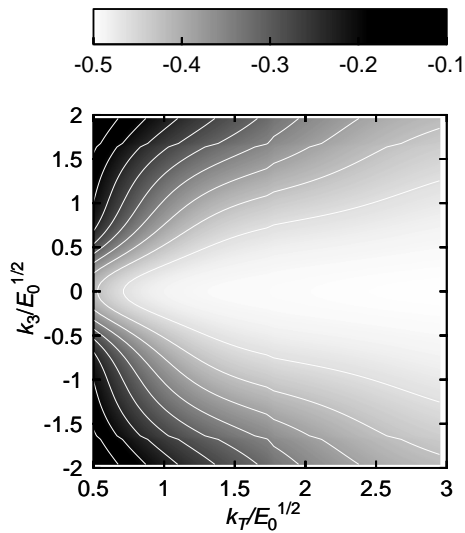}}
}
\caption{
The momentum dependence of the transverse, color singlet component
$b^s_\perp(k_\perp,k_3)$ at 
$t=2/E_0^{1/2}$ for the Bjorken expanding scenario.
}
\label{bsperp}
\end{figure}

\begin{figure}[b]
\centerline{
\rotatebox{0}{\includegraphics[height=7.0truecm]
   {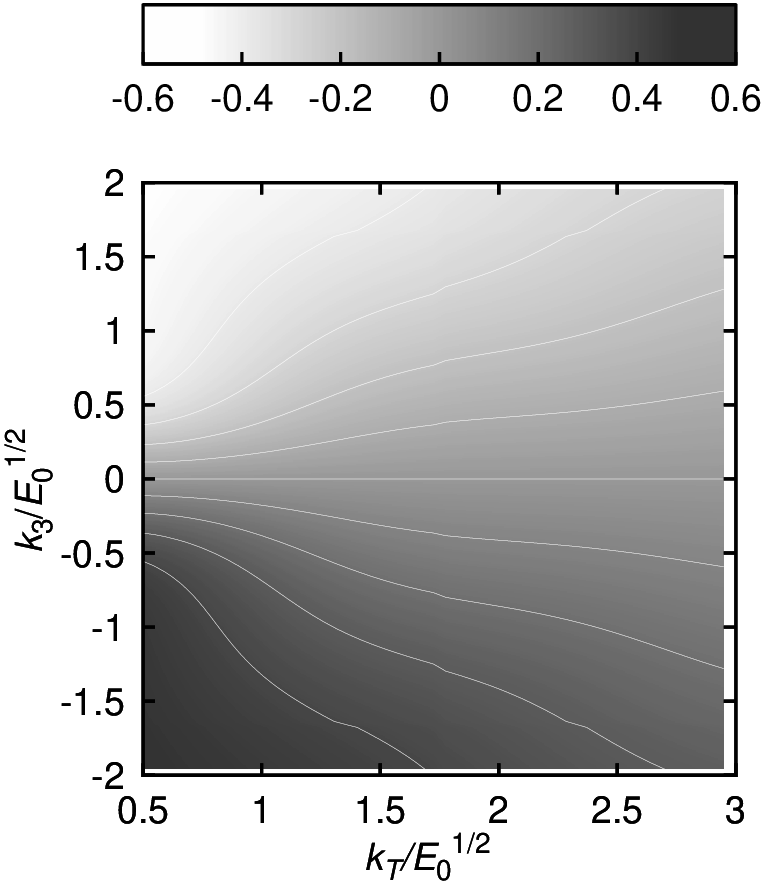}}
}
\caption{
The momentum dependence of the longitudinal, color singlet component
$b^s_3(k_\perp,k_3)$ at 
$t=2/E_0^{1/2}$ for the Bjorken expanding scenario.
}
\label{bspar}
\end{figure}

\begin{figure}[t]
\centerline{
\rotatebox{0}{\includegraphics[height=7.0truecm]
   {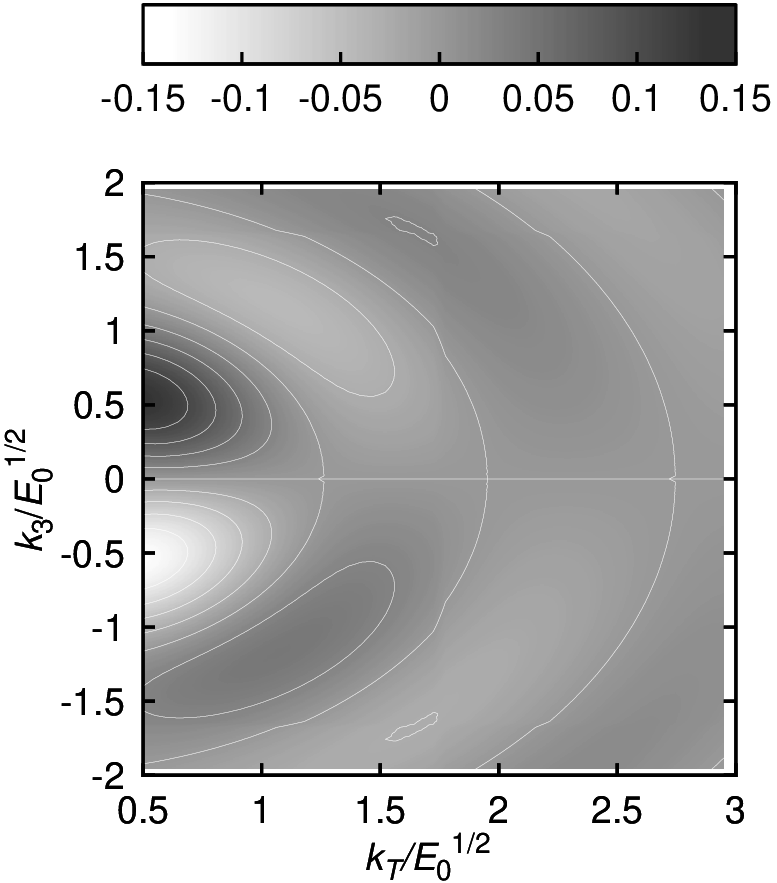}}
}
\caption{
The momentum dependence of the transverse,
color  component
$b^\diamond_\perp(k_\perp,k_3)$ at 
$t=2/E_0^{1/2}$ for the Bjorken expanding scenario.
}
\label{col_bsperp}
\end{figure}

\begin{figure}[b]
\centerline{
\rotatebox{0}{\includegraphics[height=7.0truecm]
   {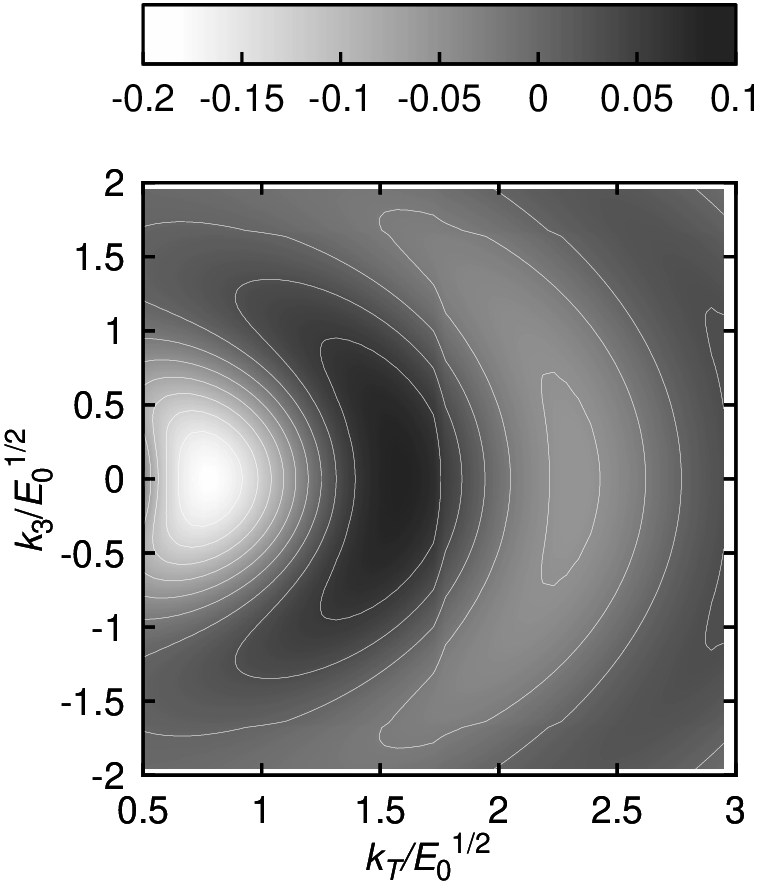}}
}
\caption{
The momentum dependence of the longitudinal, 
color component
$b^\diamond_3(k_\perp,k_3)$ at 
$t=2/E_0^{1/2}$ for the Bjorken expanding scenario.
}
\label{col_bspar}
\end{figure}

Before comparing the momentum distributions obtained numerically
in the three different physical scenarios of eqs.(\ref{Ep})-(\ref{Es})
we investigate our results on the quantities responsible for the
momentum distribution.  Detailed numerical results are shown 
in the following five figures for the specific case
of the Bjorken expansion.

The distribution function $f_f(k_\perp,k_3)$ depends directly on 
$b^s_\perp(k_\perp,k_3)$ and $b^s_3(k_\perp,k_3)$ as 
eq.~(\ref{masslesDF_fin}) shows. Figure \ref{bsperp} and \ref{bspar}
display the magnitudes of
these quantities in 2-dimensional plots. The $k_3$-symmetry
of $b^s_\perp(k_\perp,k_3)$ can be seen clearly, as well as the
asymmetric behaviour of $b^s_3(k_\perp,k_3)$. The value of these functions
are in ${\cal{O}}(1)$.

The distribution function $f_f(k_\perp,k_3)$ is color neutral,
thus the color quantities 
$b^\diamond_\perp(k_\perp,k_3)$ and $b^\diamond_3(k_\perp,k_3)$
do not contribute directly. However,
they play important role in the kinetic equation, see   
eqs.~(\ref{final_massles_iso_beg})-(\ref{final_massles_iso_end}).
Their oscillating behaviour can be seen on Figure 
\ref{col_bsperp} and \ref{col_bspar}.


\begin{figure}[t]
\centerline{
\rotatebox{0}{\includegraphics[height=7.0truecm]
   {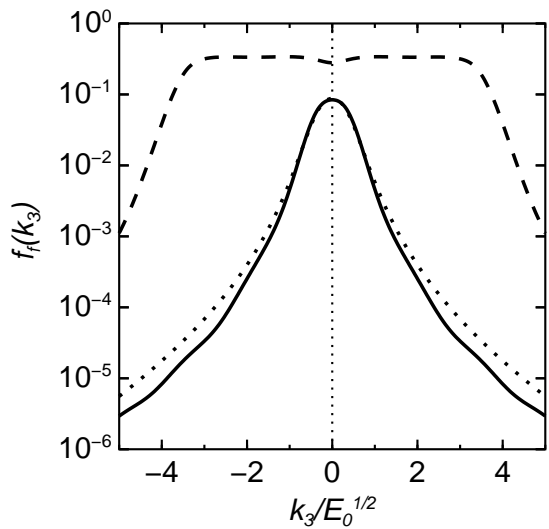}}
}
\caption{
Longitudinal momentum ($k_3$) spectra for fermions 
at $k_\perp/E_0^{1/2} = 0.5$ and $t= 2/ E_0^{1/2}$
in the three physical scenarios (see Fig. 1 and text
for explanation).
}
\label{ferpar05}
\end{figure}

\begin{figure}[b]
\centerline{
\rotatebox{0}{\includegraphics[height=7.0truecm]
   {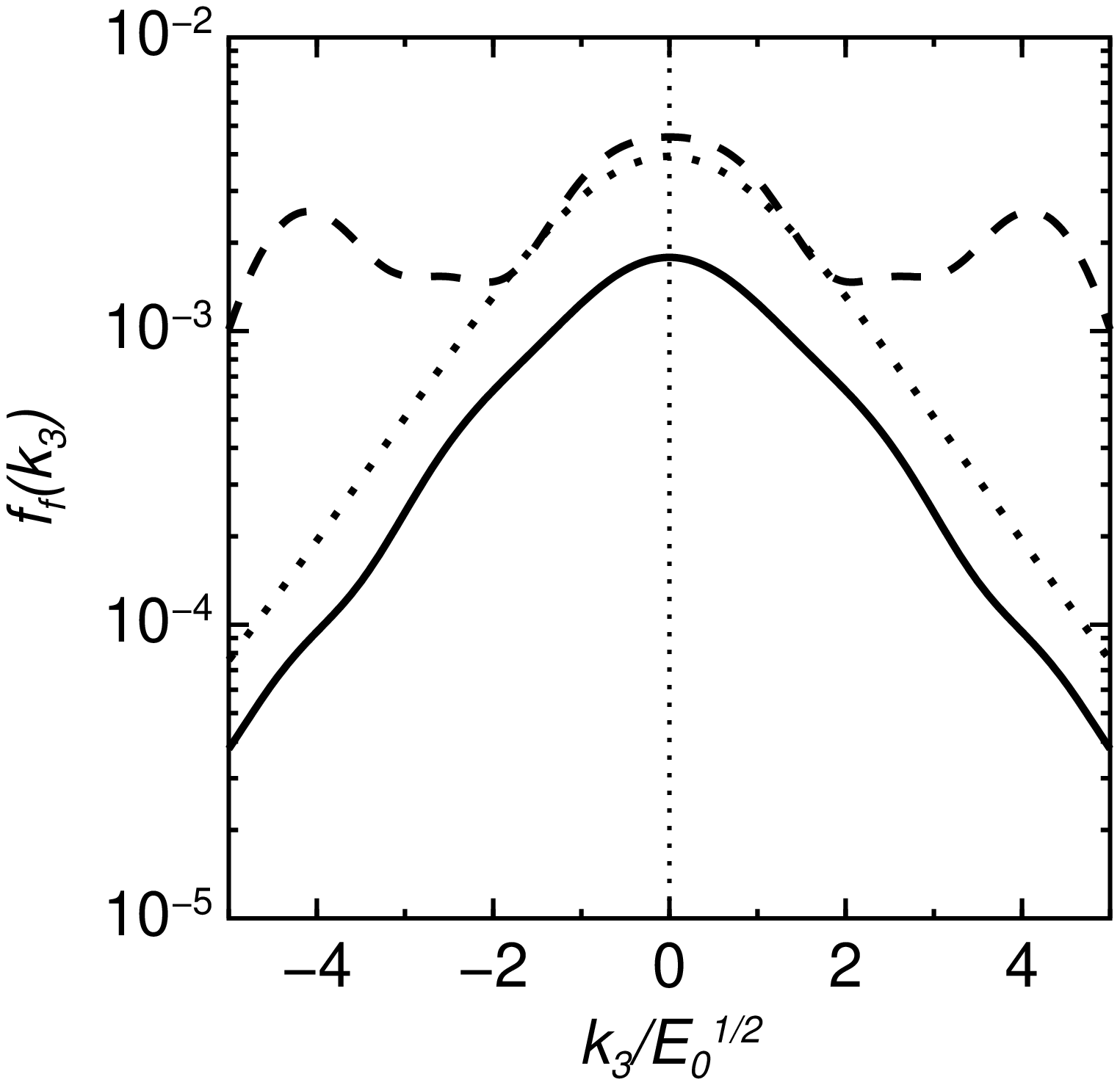}}
}
\caption{
Longitudinal momentum ($k_3$) spectra for fermions 
at $k_\perp/E_0^{1/2} = 2.5$ and $t= 2/ E_0^{1/2}$
in the three physical scenarios (see Fig. 1 and text
for explanation).
}
\label{ferpar}
\end{figure}

\begin{figure}[t]
\centerline{
\rotatebox{0}{\includegraphics[height=7.0truecm]
   {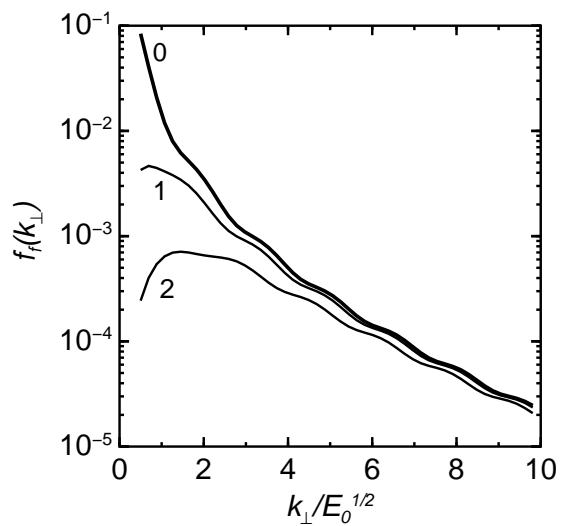}}
}
\caption{
Transverse momentum spectra for fermions 
at $k_3/E_0^{1/2} = 0, 1, 2$
in the time step $t= 2/ E_0^{1/2}$ for the Bjorken expansion scenario.
}
\label{trans_k3}
\end{figure}

Now we return to the three physical scenarios with different time dependence
described in eqs.~(\ref{Ep})-(\ref{Es}) and investigate the fermion spectra
at a large time, $t=2/E_0^{1/2}$.
Figure \ref{ferpar05} indicates the behaviour of the 
longitudinal momentum spectra at small transverse momentum value,
where we choose $k_\perp/E_0^{1/2}=0.5$. 
Pulse-type time dependence leads to a narrow $k_3$-distribution 
({\it dotted line}),
which mimics a Landau-type hydrodynamical initial condition.
The longitudinal spectra from constant field scenario
({\it dashed line}) leads to a flat distribution function
in $k_3$. This result agrees well with a 1-dimensional,
longitudinally invariant hydrodynamical initial
condition, as we expect. 
Considering the scaled field scenario ({\it solid line}), since its time
dependence is very similar to the pulse-type case (see Fig.~\ref{etime}), 
then the similarity in the longitudinal spectra
is well understandable.

If we increase the transverse momentum and choose
$k_\perp/E_0^{1/2}=2.5$, then Figure \ref{ferpar} displays the obtained
longitudinal spectra:
the $k_3$ dependence becomes very
similar in the small $k_3$ region for the three different time evolution.
For pulse-type time dependence ({\it dotted line}) 
and for constant field scenario ({\it dashed line}) even the magnitude 
is very close to each other. Further numerical investigations are needed
to understand this similarities.

The interplay between transverse and longitudinal momentum spectra is 
displayed on Figure \ref{trans_k3}.  In the Bjorken expansion
scenario of eq.~(\ref{Es}) we choose different longitudinal momentum windows,
namely $k_3/E_0^{1/2}=0, 1, 2$ and extract the
transverse momentum spectra at $t= 2/ E_0^{1/2}$.
On Figure \ref{trans_k3} one can see large deviation at small transverse 
momenta and similar spectra at high transverse momenta: 
hard fermion production is similar at different rapidities, but soft 
production (and any 'effective temperature') is very strongly 
rapidity dependent.

\begin{figure}[t]
\centerline{
\rotatebox{0}{\includegraphics[height=7.0truecm]
   {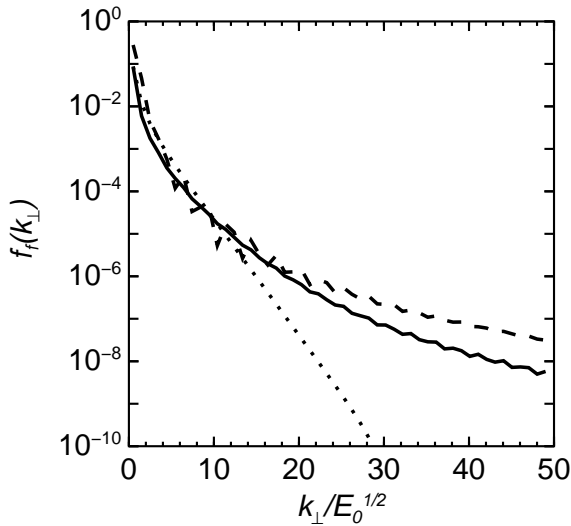}}
}
\caption{
The transverse momentum spectra for fermions 
at $k_3=0$ and $t= 2/ E_0^{1/2}$
in the three physical
scenarios (see Fig.~\ref{etime} and text for explanation).
}
\label{ferperp}
\end{figure}

Figure \ref{ferperp} displays the transverse momentum spectra 
for the three different physical scenario at momentum $k_3=0$
and  time $t=2/E_0^{1/2}$.
Pulse-type time dependence leads to exponential spectra ({\it dotted line}),
$f_f \propto \exp (k_T/T)$ with
slope value $T = 1.54 \cdot  E_0^{1/2}$. 
In the other two cases,  we obtain non-exponential spectra
generated by the long-lived field. 
Here the spectra from constant 
({\it dashed lines}) and scaled ({\it solid lines}) fields are close,
because the production and annihilation rates 
balance each other.  Slight differences appear because
of the fast fall of the scaled field immediately after $t = 0$.


\begin{figure}[b]
\centerline{
\rotatebox{0}{\includegraphics[height=7.0truecm]
   {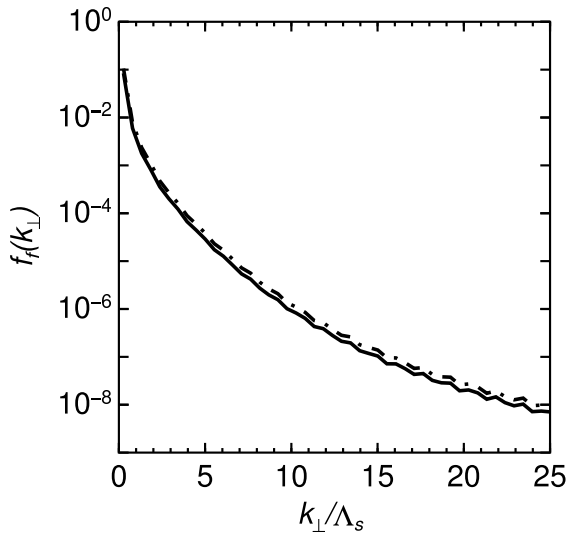}}
}
\caption{
Transverse momentum spectra of fermions from our
calculation with scaled time evolution, $E_{scaled}(t)$,
for SU(2) case (solid line), and from Ref.~\cite{SkokLev05}
for U(1) (dash-dotted line).
}
\label{su2_u1}
\end{figure}

\begin{figure}[b]
\centerline{
\rotatebox{0}{\includegraphics[height=7.0truecm]
   {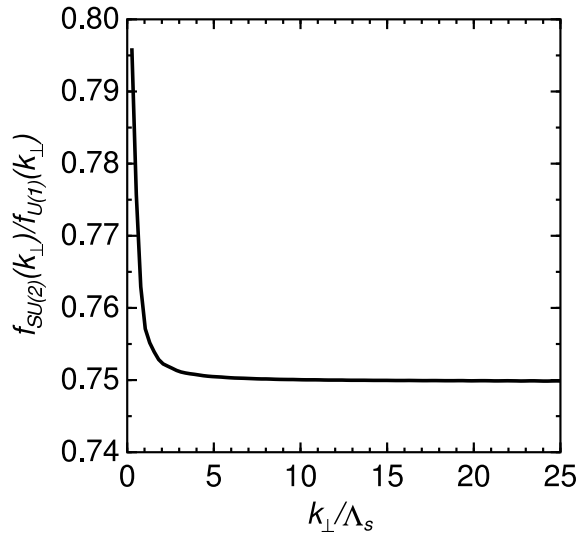}}
}
\caption{
The ratio of the numerical results from Figure~\ref{su2_u1} on
transverse momentum spectra for fermions in a linear scale:
the SU(2) case (this calculation)
is divided by the U(1) case from Ref.~\cite{SkokLev05}. }
\label{rat_su2_u1}
\end{figure}

In our previous paper~\cite{SkokLev05} we considered fermion production 
in U(1) classical field. Comparison between current SU(2) calculation
and U(1) results can be made, if we fix both field strength at the same value. 
Taking the value  $E_0^{1/2}/\Lambda_s = 0.54$ used in Ref.~\cite{SkokLev05}, 
we obtain transverse momentum spectra depicted in the Figure~\ref{su2_u1}, 
where particle production in the scaled field scenario was considered. 
 We have found that the obtained transverse distribution functions 
for fermions
are close to each other and have at almost the same shape. 
Figure~\ref{rat_su2_u1} displays the ratio of the two numerical results
on a linear scale.
It clearly shows the  appearance of a numerical value of
$f_{\rm SU(2)}(k_\perp)/f_{\rm U(1)}(k_\perp)=0.75$ for transverse momenta 
$k_\perp\geq\Lambda_s$.
This value may indicate the presence of scaling
solutions of the investigated kinetic equations. 
The direct study of the kinetic equations
in Section IV and V does not reveal scaling, but special cases can
indicate such a behaviour.

\section{Scaling solutions }

In this section we consider specific initial conditions and corresponding solution for
the kinetic equations eqs.~(\ref{final_massles_iso_beg})-(\ref{final_massles_iso_end}).
At first we assume
the singlet and the color multiplet components have the same initial values. 
This ``symmetry'' is generally violated by vacuum initial conditions demanding 
$b^\diamond$ and $d^\diamond$ to be zero in vacuum in contract to initial conditions for 
$b^s$ and $d^s$ (cf. (\ref{initial_cond_beg})-(\ref{initial_cond})). 
Accepting this assumption, one may note that  
eqs.~(\ref{final_massles_iso_beg})-(\ref{final_massles_iso_end})
have the following special solution: 
$b^\diamond = \eta b^s$ and $d^\diamond=\eta d^s$, 
where $\eta$ is a numerical value to be defined 
in Appendix A. 
In this case the SU(2) kinetic
equations for massless fermions can be shortened and
rewritten into the following form (see Appendix A for details):
\begin{eqnarray}
\dot{f}  &=& \frac{1}{2} W v \ , \label{abel_KE1} \\
\dot{v}  &=& W (1 - 2 f) - 2 \vert \mathbf{k} \vert u \ , \label{abel_KE2}\\
\dot{u}  &=& 2 \vert \mathbf{k} \vert  v \ . 
\label{abel_KE3}
\end{eqnarray}
One can recognize that formally these equations have been used 
earlier in the Abelian case to calculate massless fermion production 
(see eqs.(22)-(24) with zero fermion mass in Ref.~\cite{SkokLev05}).

The region of our interest is the small longitudinal ($k_3\to0$) and
high transverse ($k_\perp\to\infty$) momenta. It is reasonably to assume
that for this region the distribution function is much smaller then unity,
$f(k_\perp\to\infty)\ll1$, demonstrating small particle yield in
the high transverse momentum region. Thus the term of $(1-2f)$
in eq.~(\ref{abel_KE2}) can be simplified to unity.

The eqs.~(\ref{abel_KE1}-\ref{abel_KE3}) allow us to introduce
new scaled variables, namely  $\hat{W}={W} \eta$, 
$\hat{f} = f \eta^2$, $\hat{v}=v \eta$ and  $\hat{u}=u \eta$.
Thus the above equations can be rewritten into the following form:
\begin{eqnarray}
\dot{\hat{f}}&=&\frac{1}{2} \hat{W} \hat{v}, \\
\dot{\hat{v}}&=&\hat{W} - 2 |\mathbf{k}| \hat{u}, \\
\dot{\hat{u}}&=&2 |\mathbf{k}| \hat{v}.
\end{eqnarray}
In the U(1) case we have solved this set of equations with the scale
parameters $\eta=1$, and obtained the $f(\eta=1)$ momentum
distribution functions~\cite{SkokLev05}.
Thus we know any scaled solution: 
$f(\eta) = {f(\eta=1)}\cdot {\eta^{-2}}$.
In the case of SU(2) we have scale variable $\eta=\pm 2/\sqrt{3}$
(see Appendix A). Thus we can easily extract the wanted ratio: 
$f_{\rm SU(2)}/f_{\rm U(1)}=3/4$. This number agrees with the
numerically calculated value in the high transverse momentum region,
as Figure~\ref{rat_su2_u1} displays.

For small transverse momentum
our analysis cannot be performed, because Pauli suppression
factor $(1-2 f)$  differs from unity (see e.g. Figure~\ref{ferpar05}). 
The numerical calculations have shown a ratio of
$f_{\rm SU(2)}/f_{\rm U(1)} > 3/4$ in this momentum region
(see Figure~\ref{rat_su2_u1}).

The physical meaning of this $\eta$-scaling is evident:
number of particles with high transverse momentum
is proportional to the second power of the field strength, 
namely $(g E)^2$.

An additional scaling property has been used during
the presentation of our numerical values. Namely in the  
eqs.~(\ref{partII_r1})-(\ref{partII_r4})
one can introduce a basic scaling in the variables:
\begin{eqnarray}
{\vec k} &\longrightarrow& \lambda \cdot {\vec k} \ ,  \\
E_0 &\longrightarrow& \lambda^2 \cdot E_0 \ ,  \\
t &\longrightarrow& \lambda^{-1} \cdot t  \ .
\end{eqnarray} 
In this case the obtained distribution function 
will not change.

Furthermore, in the case of external pulse field of eq.~(\ref{Ep}),
at $t \longrightarrow \infty$ the exponential transverse spectra of 
the stationary solution for the distribution function $f$
can be characterized by an effective temperature as we have shown
on Figure~\ref{ferperp}.
This temperature will be scaled as the momentum,
$T \longrightarrow \lambda \cdot T$.

In our previous paper~\cite{SkokLev05} for the Abelian external field
with scaled time evolution
we have obtained numerical transverse spectra, which was very
close to the perturbative QCD results for high-$k_T$,
namely scaling with 
$\log(k_T/\Lambda_s)\cdot (k_T/\Lambda_s)^4$~\cite{Gyul97}.
As Figures~\ref{su2_u1} and ~\ref{rat_su2_u1} displayed,
we obtain the same behaviour for the SU(2) case,
although we can not prove this scaling analytically.

\section{Conclusion}

In this paper we investigated fermion  production from
a strong classical SU(2) field in a kinetic model.
We derived the appropriate system of differential equations, 
which became simplified after introducing zero fermion mass. 
Following the physical ideas suggested in our earlier paper
on the U(1) case (see. Ref.~\cite{SkokLev05}), 
we assumed three different time dependence for the  longitudinal 
color field of fixed direction in color space  and obtained three different types of longitudinal and transverse 
momentum spectra for the  SU(2) fermions. We have found 
that the time dependence of the external field determines 
if the transverse momentum spectra are exponential or polynomial, 
similarly to the U(1) case.
We solved the kinetic equation
numerically and displayed the obtained results on longitudinal and
transverse momentum distributions for the massless fermions.
Furthermore, for the scaled time dependent external field 
in the transverse momentum spectra
we obtained a constant ratio of 0.75 at high $p_T$
between the recent SU(2) and earlier U(1) results.
We could reproduce this scaling behaviour analytically
in the high transverse momentum limit, together with the factor of 0.75.
Furthermore, our numerical results display
the presence of a 
$\log(k_T/\Lambda_s)\cdot (k_T/\Lambda_s)^4$ scaling,
similar to results obtained from
perturbative QCD calculations.

\section{Appendix A}

In this Appendix we display a specific solution of the SU(2)
kinetic equation of (\ref{wigner_gen}), which leads to the form
of equations
known and solved already in the U(1) case (see Ref.~\cite{SkokLev05}).
These equations allow us to recognize scaling properties
in U(1) and SU(2), discussed in Section VII.

Considering the form of SU(2)
eqs.~(\ref{final_massles_iso_beg})-(\ref{final_massles_iso_end}),
they can be solved assuming a very specific condition for the singlet 
and the color multiplet components~\cite{Prozor03}, namely
\begin{eqnarray}
b^\diamond = \eta b^s \ ,  \label{diam1} \\
d^\diamond = \eta d^s \ .  \label{diam2}
\end{eqnarray}
Substituting this constraint into 
eqs.~(\ref{final_massles_iso_beg})-(\ref{final_massles_iso_end}),
a definite value can be obtained for parameter $\eta$,
namely $\eta=\pm 2 /\sqrt{3}$. 

In parallel, these equations become even simpler:
\begin{eqnarray}
\label{final_massles_isotrop_beg}
(\partial_t + \frac{g}{\eta} E^\diamond
\frac{\partial}{\partial k_3}) b^s_\perp &=& -2 k_3 d^s, \\
(\partial_t + \frac{g}{\eta} E^\diamond \frac{\partial}{\partial k_3}) b^s_3
&=& 2 k_\perp d^s, \\
(\partial_t + \frac{g}{\eta} E^\diamond \frac{\partial}{\partial k_3}) d^s
&=& 2 k_3 b^s_\perp - 2  k_\perp b^s_3;
\label{final_massles_isotrop_end}
\end{eqnarray}

For further consideration it is appropriate to rewrite this set of equations in terms
of the distribution function $f_f(\vec{k},t)$. Applying the differential operator 
\begin{equation}
D_t = \partial_t + \frac{g}{\eta} E^\diamond \frac{\partial}{\partial k_3}
\end{equation}
to the distribution function (\ref{masslesDF_fin})  and using
(\ref{final_massles_isotrop_beg})-(\ref{final_massles_isotrop_end}) we obtain
\begin{eqnarray}
D_t f &=& \frac{1}{2} W v, \\
D_t v &=& W (1-2f) -2 |\vec{k}| u, \\
D_t u &=& 2 |\vec{k}| v;
\end{eqnarray}
where we introduced the functions $W,v,u$ as
\begin{eqnarray}
W &=& \frac{g}{\eta} \frac {E k_\perp} {\mathbf{k}^2 } \\
v &=& 2\frac{k_\perp}{\vert \mathbf{k} \vert} b^s_3 -
2\frac{k_3}{\vert \mathbf{k} \vert} b^s_\perp, \\
u &=& -2 d^s.
\label{mapping}
\end{eqnarray}

Using method  of characteristics~\cite{courant} we finally get:
\begin{eqnarray}
\dot{f} &=& \frac{1}{2} W v, \\
\dot{v} &=& W (1-2f) -2 |\vec{k}| u, \\
\dot{u} &=& 2 |\vec{k}| v;
\end{eqnarray}
where $k_3$ becomes time-dependent
\begin{equation}
\dot{k}_3 = \frac{g}{\eta} E^\diamond.
\label{k_3}
\end{equation}

One can recognize that the same equations have been used
to calculate fermion production in the Abelian case with $\eta=1$ 
(see eqs.(22)-(24) with zero fermion mass in Ref.~\cite{SkokLev05}).
This surprising discovery indicates that in very specific cases
Abelian-like equations can be derived from non-Abelian kinetic equations.

However, there is a conflict between the constraint in 
eqs.(\ref{diam1})-(\ref{diam2}) and the physical vacuum solution
in eq.~(\ref{WFvac}), which indicates a more complicated connection
between the non-Abelian and Abelian cases.

\section*{ Acknowledgments }

 We thank Yu.B. Ivanov  for stimulating discussions.
This work was supported in part by  Hungarian grant OTKA-T043455,
NK062044, IN71374, the MTA-JINR Grant, 
RFBR grant No. 05-02-17695 and a special  program of 
the Ministry of Education and Science of the Russian Federation, 
grant RNP 2.1.1.5409.

\end{document}